\documentclass{ws-procs975x65}

\usepackage{epsfig}
\usepackage{epstopdf}
\usepackage{amsmath}
\usepackage{amssymb}
\usepackage{enumerate}
\usepackage{relsize}
\usepackage{xcolor}
\usepackage{graphicx}

\renewcommand\vec[1]{\boldsymbol{#1}}

\newcommand{\grad}{\vec{\nabla}}
\newcommand{\gradx}{\grad_{\!\!x}}

\newcommand{\gradq}{\grad_{\!\!q}}
\newcommand{\nablax}{\nabla_{\!\!x}}
\newcommand{\D}{D_{\!\mathsmaller{\mathsmaller{+} } }}

\begin{document}

\title{\uppercase{Adhesion and the Geometry of the Cosmic Web}}

\author{\uppercase{Johan Hidding}$^*$, \uppercase{Rien van de Weygaert},
\uppercase{Gert Vegter} \\ and \uppercase{Bernard J.T. Jones}}

\address{Kapteyn Astronomical Institute, University of Groningen,\\
Groningen, 9747 AD, The Netherlands\\
$^*$E-mail: hidding@astro.rug.nl\\
www.rug.nl/sterrenkunde}

\begin{abstract}
	We present a new way to formulate the geometry of the Cosmic Web
	in terms of Lagrangian space. The Adhesion model has an
	ingenious geometric interpretation out of which the spine of the
	Cosmic Web emerges naturally. Within this context we demonstrate
	a deep connection of the relation between Eulerian and Lagrangian
	space with that between Voronoi and Delaunay tessellations.
\end{abstract}

\keywords{Cosmology: large-scale structure of universe; cosmic web;
adhesion model}

\bodymatter

\section{Introduction}
The Cosmic Web is the largest known structure in the Universe. It is
seen in both observations \cite{deLapparent1986} and simulations
\cite{Springel2005}. The Adhesion model
\cite{Gurbatov1984,Shandarin1989,Vergassola1994} provides a heuristic
model describing both the intricate geometry of the Cosmic Web, and the
non-local nature of its dynamics. Before shell crossing occurs, Adhesion
follows the Zel'dovich approximation \cite{Zeldovich1970}
\begin{equation}
\vec{x}(\vec{q}, \D) = \vec{q} - \D \gradq \Phi_0(\vec{q}),
\end{equation}
describing the motion of a particle labeled $\vec{q}$ in 
comoving coordinates as having a constant comoving velocity $\vec{v} =
\partial \vec{x} / \partial \D = - \gradq \Phi$. The
Adhesion model adds a viscosity term to this recipe, emulating the
adhesive effects of gravity, preventing shell-crossing from ever 
taking place. The resulting equation of motion is known as Burgers' equation
\begin{equation}
\frac{\partial \vec{v}}{\partial \D} + \left(\vec{v} \cdot
\gradx\right) \vec{v} = \nu \nablax^2 \vec{v}.
\end{equation}
In the limit where $\nu \to 0$, it has the exact solution \cite{Hopf1950}
\begin{equation}
\Phi(\vec{x}, \D) = \max_q \left[ \Phi_0(\vec{q}) - \frac{(\vec{x} -
\vec{q})^2}{2\D}\right].
\label{eq:burgsol}
\end{equation}
This solution gives us not only the potential at some growing-mode time $\D$,
but also the mapping from Eulerian coordinates $\vec{x} \in \mathcal{E}$ to Lagrangian
coordinates $\vec{q} \in \mathcal{L}$, as the Lagrangian coordinate where the maximum is
attained. 

\section{Weighted Voronoi diagrams}
We found expression (\ref{eq:burgsol}) to be identical to that of the \emph{weighted
Voronoi diagram} of the set of points $\vec{q}$, weighted by the value
of the velocity potential \cite{Hidding2012}.  The weighted Voronoi cell
of a point $\vec{q} \in \mathcal{L}$ is given by 
\begin{equation}
V_q = \left\{\vec{x} \in \mathcal{E}\ \Big|\ (\vec{x} - \vec{q})^2 + w_q\leq
(\vec{x} - \vec{p})^2 + w_p,\ \vec{p} \in \mathcal{L}\right\}.
\end{equation}
The physical interpretation of this expression is that the Voronoi cell
gives us the Eulerian region of space a parcel of matter occupies. If we
invert this relation, we find that the Delaunay triangulation, being the
dual of the Voronoi diagram tells us where the matter around
an Eulerian location came from. Describing cosmic structures in
this way, allows us to trace their formation and change in topology over
time.


\section{Evolution of the Cosmic Web}
\begin{figure}[b]
	\centering
	\begin{minipage}[c]{\textwidth}
	\includegraphics[width=0.49\textwidth]{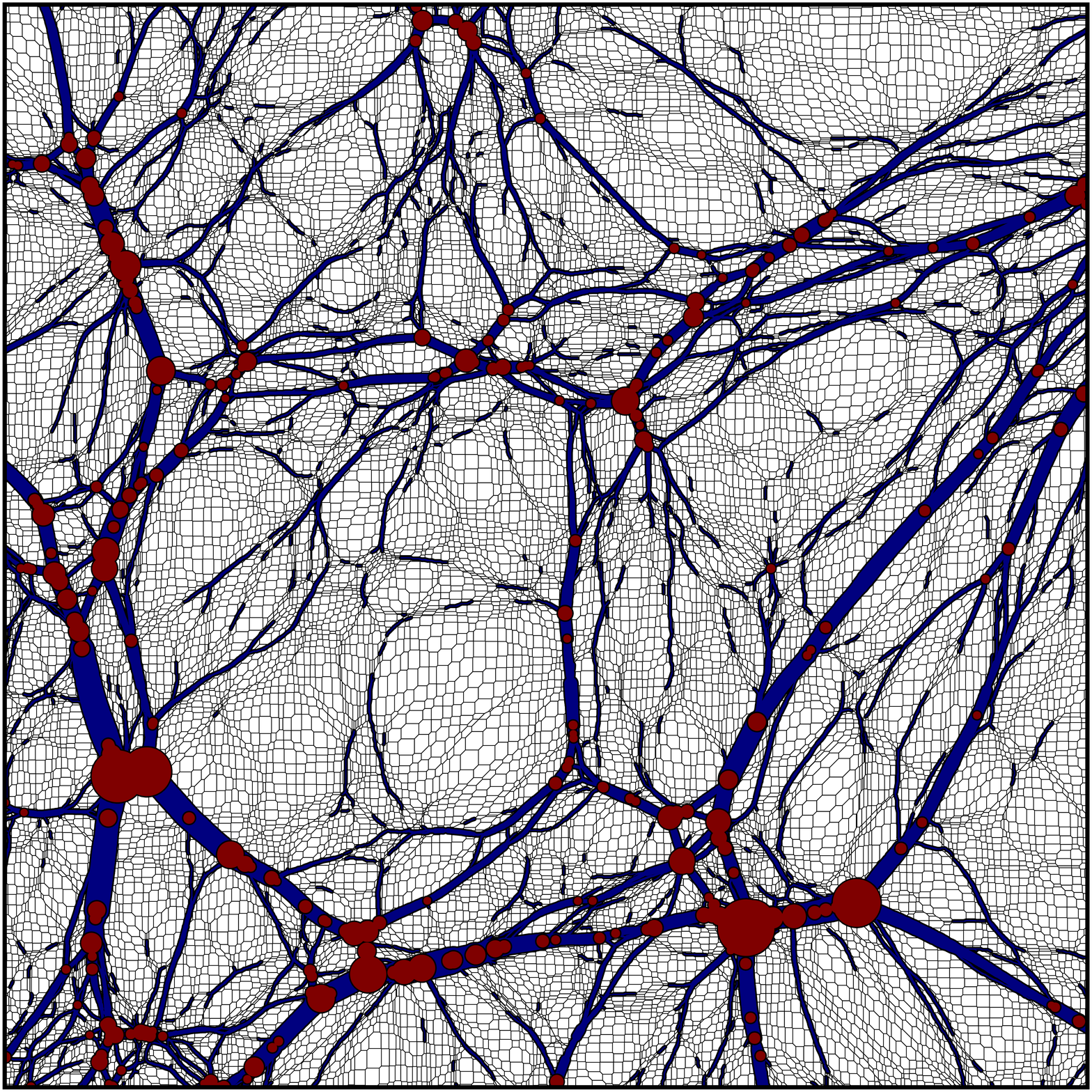}
	\includegraphics[width=0.49\textwidth]{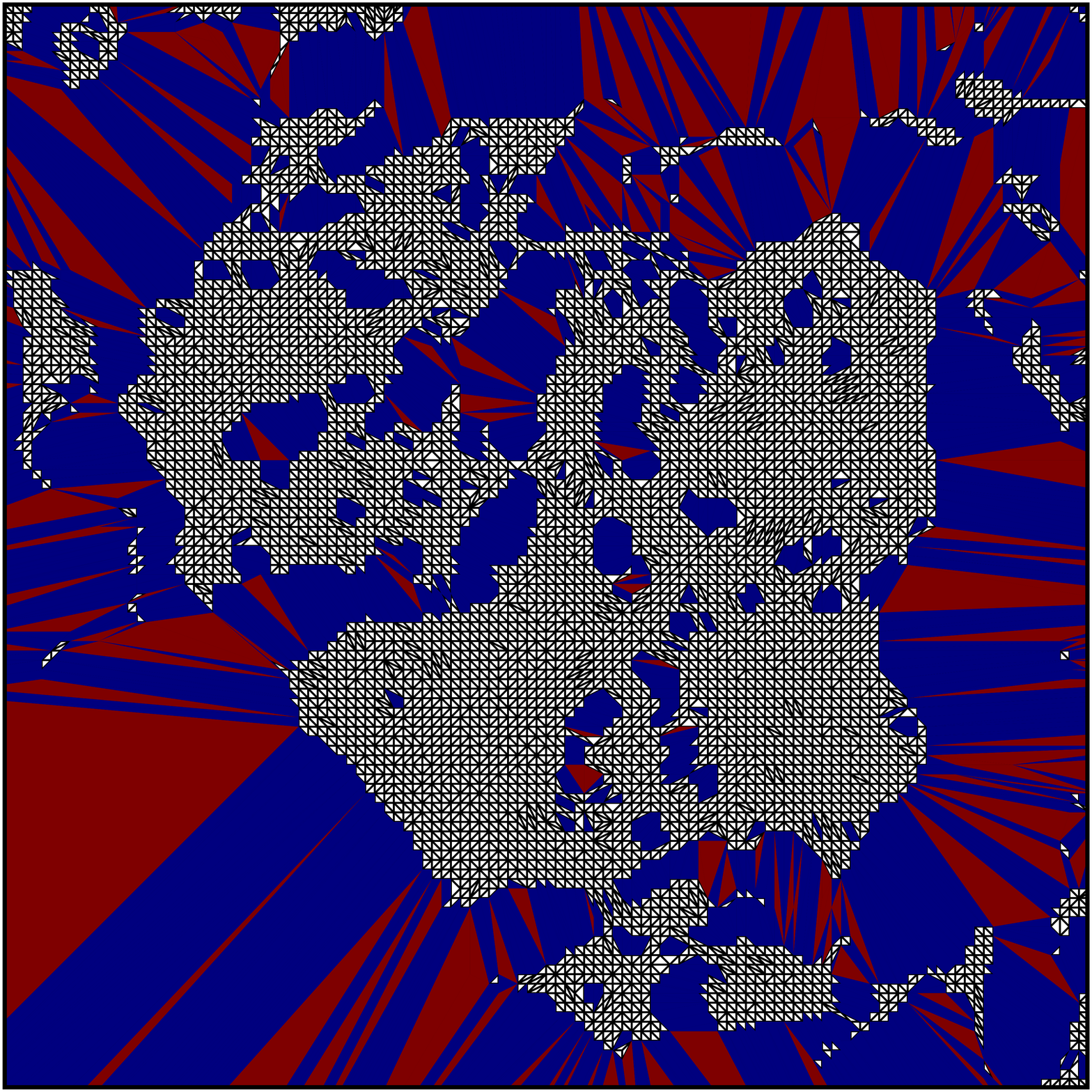}
	\end{minipage} \\\vspace{9pt}
	\begin{minipage}[c]{\textwidth}
	\includegraphics[width=0.49\textwidth]{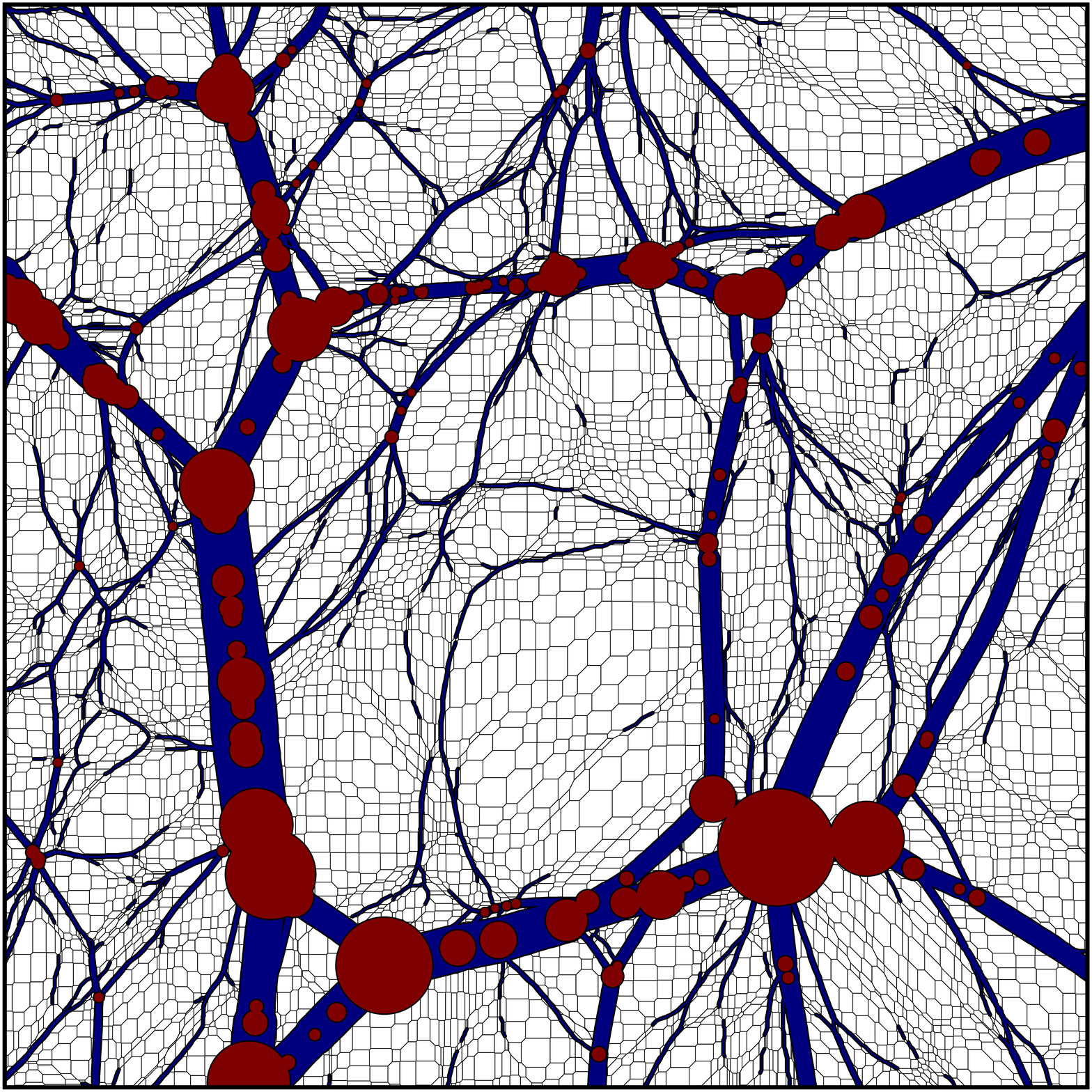}
	\includegraphics[width=0.49\textwidth]{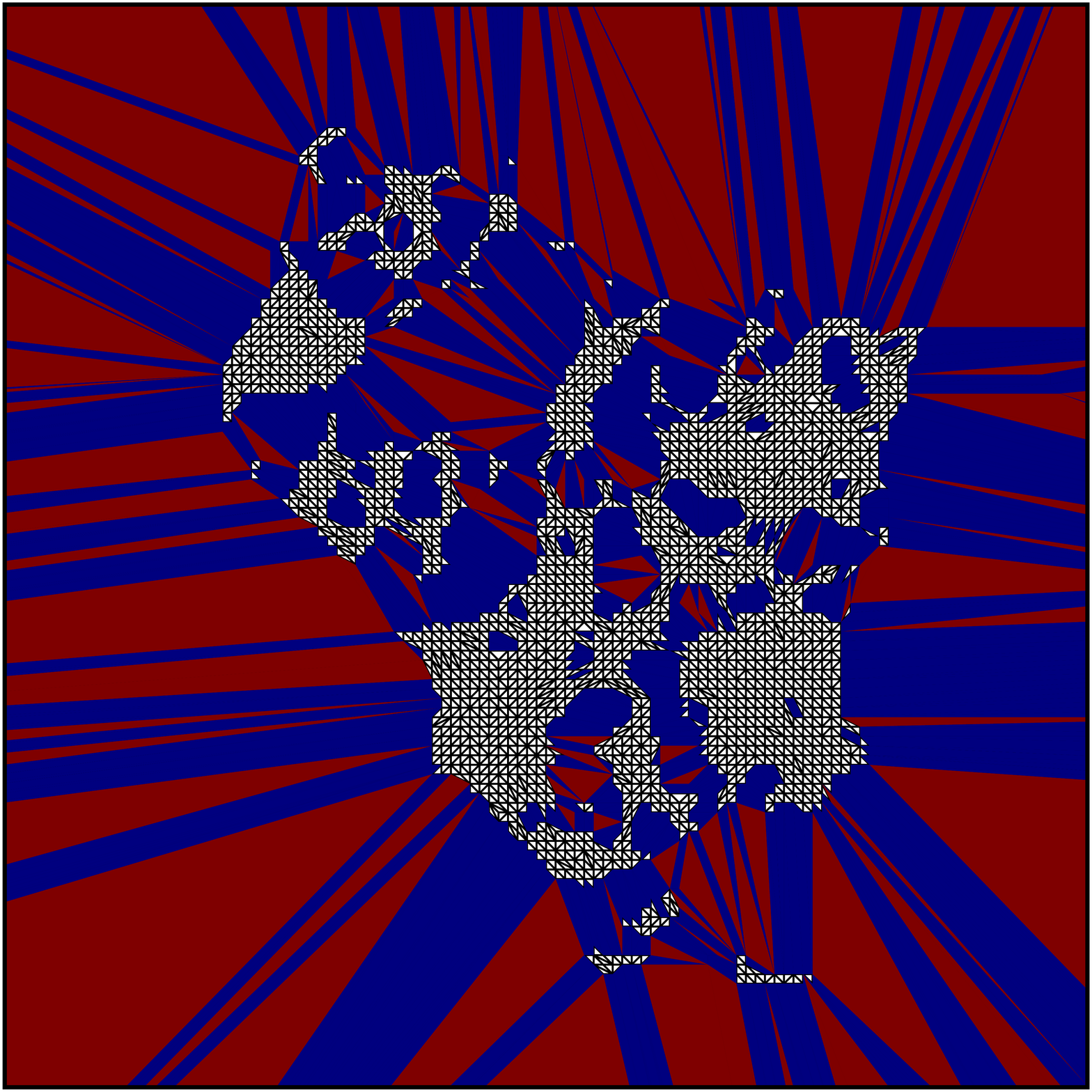}
	\end{minipage} \\\vspace{9pt}
	\begin{minipage}[c]{\textwidth}
	\includegraphics[width=0.49\textwidth]{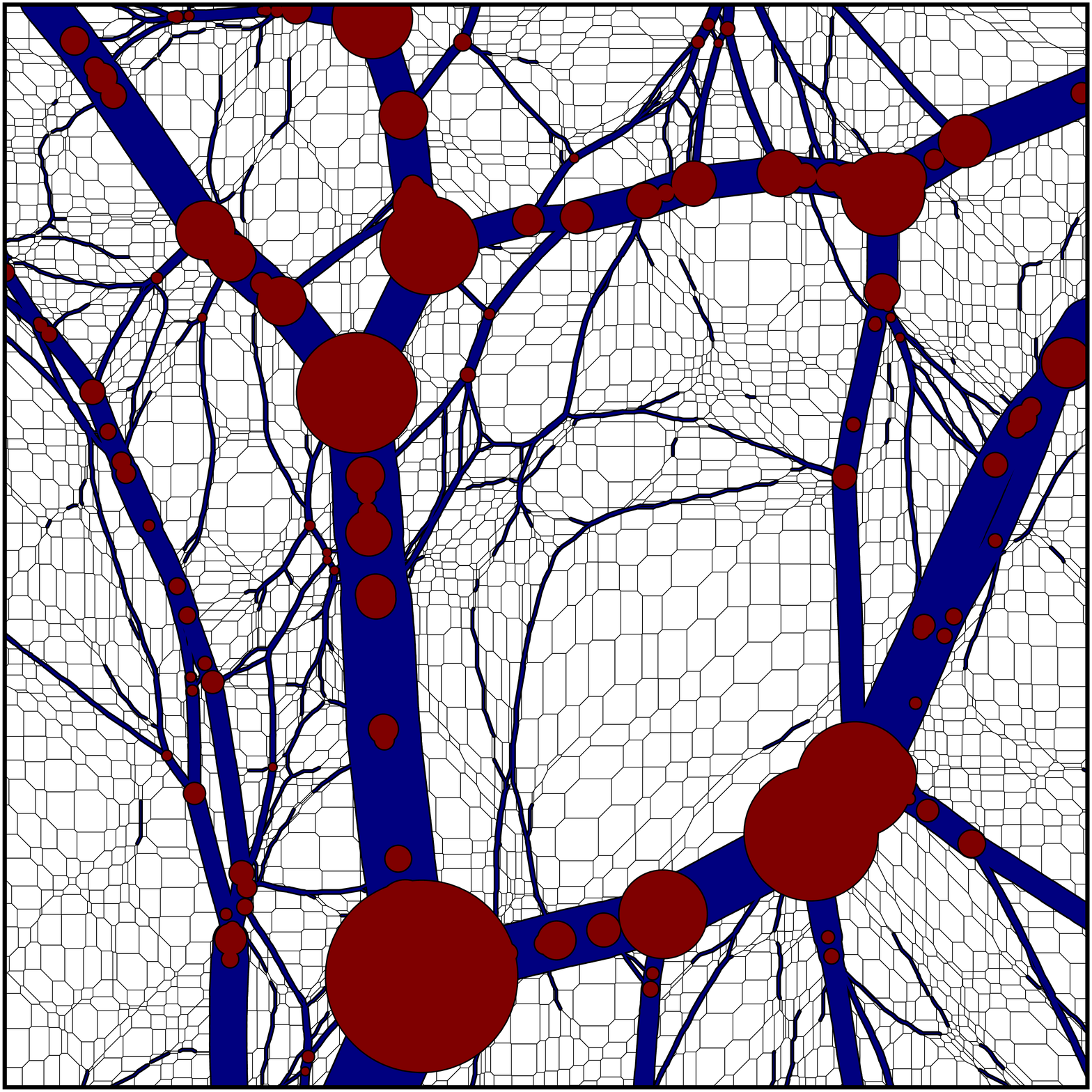}
	\includegraphics[width=0.49\textwidth]{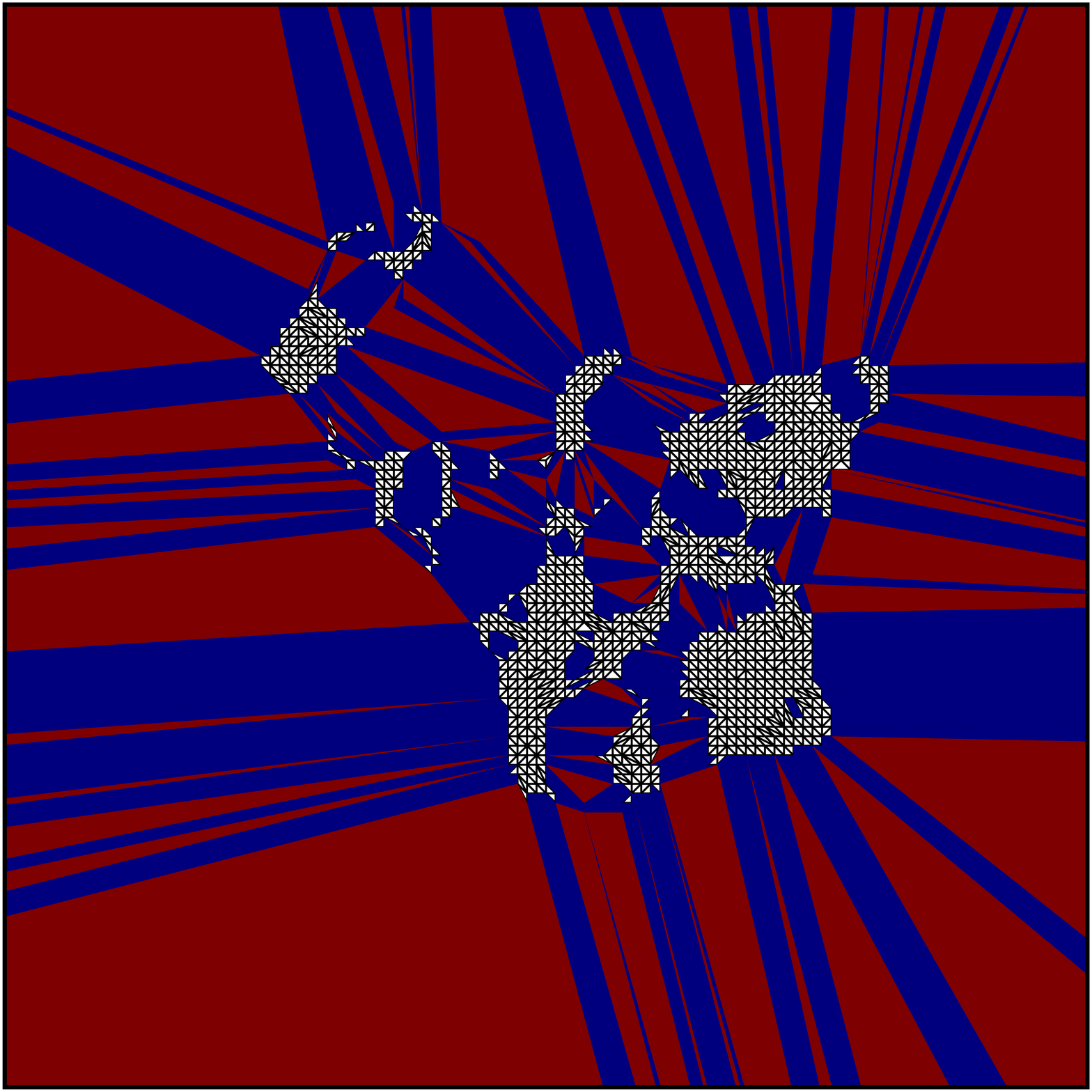}
	\end{minipage}
	\caption{Time evolution in Eulerian (left) and Lagrangian
	(right) space of a 2-D model. On the right we see the Lagrangian areas 
	corresponding to the clusters (red triangles), and filaments (blue
	lenticular regions) shown on the left. White is for voids.}
	\label{fig:lnstev}
\end{figure}

We can find the walls of the web structure as edge-like objects in
Lagrangian space, whereas filaments have a flattened signature. Clusters,
being the most massive concentrations of matter are therefore most
extended in Lagrangian space. Looking at figure \ref{fig:lnstev}, we
see voids growing to compress their weaker neighbours, while their area
in Lagrangian space shrinks as matter flows into the walls and filaments
bounding the voids.

\begin{figure}[t]
	\centering
	\includegraphics[width=0.32\textwidth]{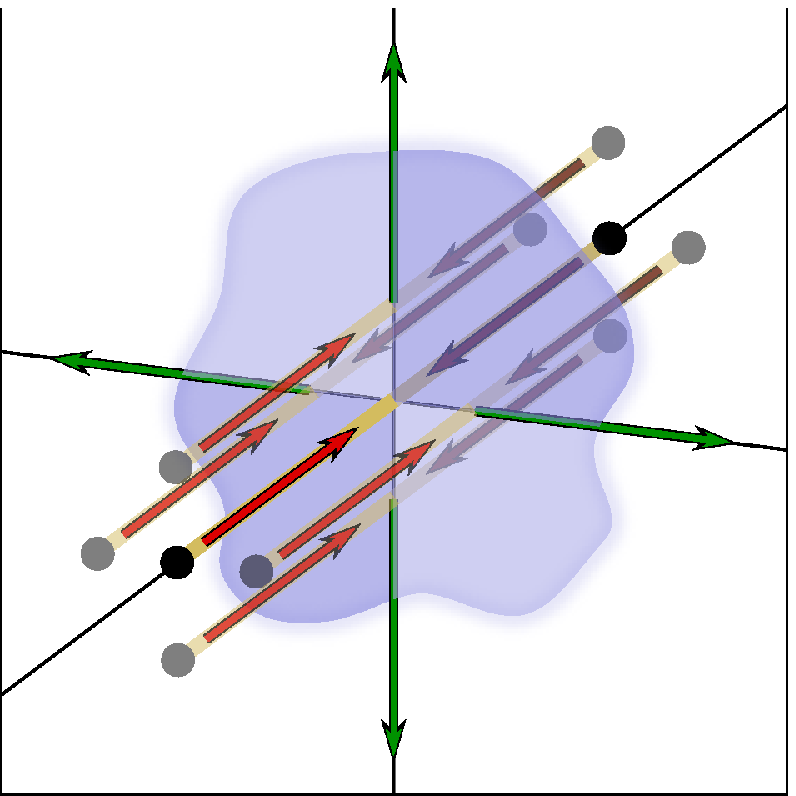}\hspace{1pt}
	\includegraphics[width=0.32\textwidth]{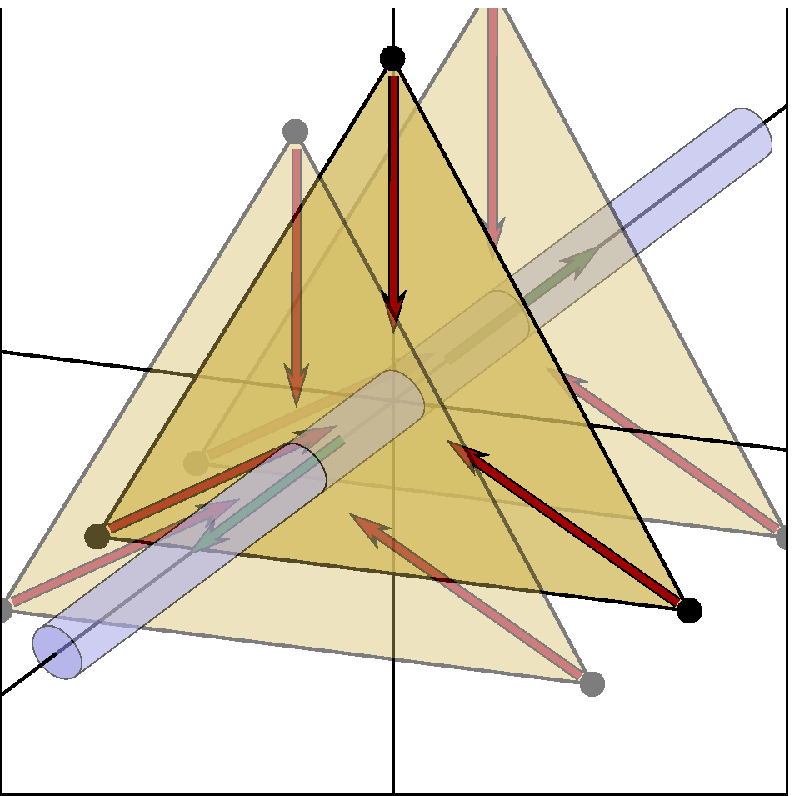}\hspace{1pt}
	\includegraphics[width=0.32\textwidth]{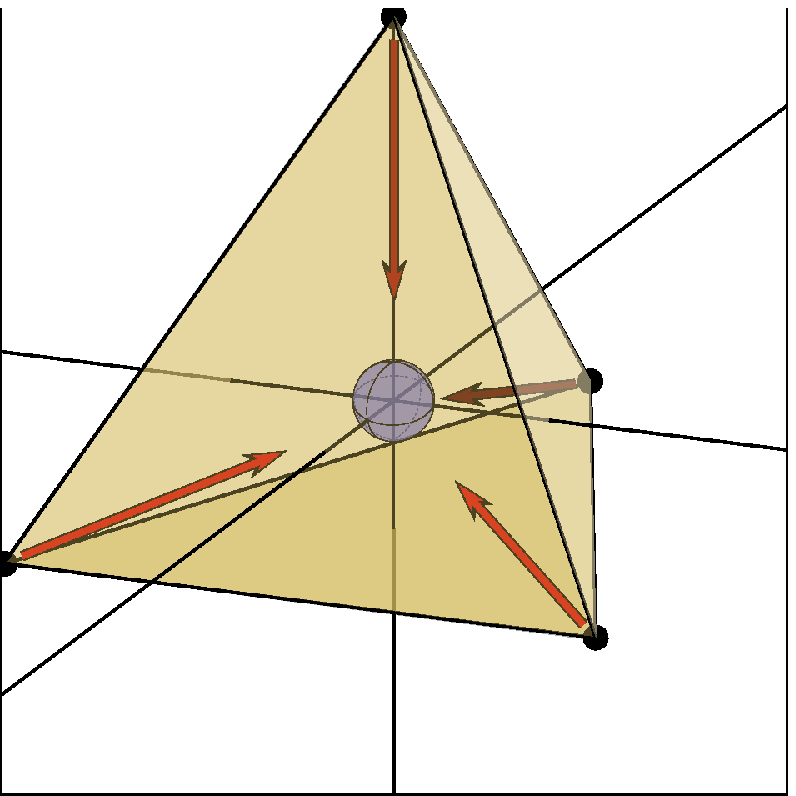}
	\caption{The duality of three morphological types: walls,
filaments and clusters. Each time the Eulerian structure is
shown in blue, and the Lagrangian mass building the structure in
yellow. The green arrows show the motion internal to the structure.
A wall is a flattened structure, where locally the mass
originates from a line-like region. A filament is elongated, however
each part of the filament was originally distributed as a sheet. The
cluster has collapsed from three directions.}
\end{figure}
	
\bibliographystyle{ws-procs975x65}

\end{document}